\documentstyle[prb,aps]{revtex}
\begin{document}
\draft

\title{Coherent Resonant Tunneling Through an Artificial Molecule}
\author{C.~A.~Stafford}
\address{Fakult\"at f\"ur Physik, Albert-Ludwigs-Universit\"at,
Hermann-Herder-Str.\ 3, D-79104 Freiburg, Germany}
\author{R. Kotlyar and S.~Das Sarma}
\address{Department of Physics, 
University of Maryland, College Park, Maryland 20742}
\date{20 January 1998}
\maketitle

\begin{abstract}
Coherent resonant tunneling through an array of quantum dots 
in an inhomogeneous
magnetic field is investigated using an extended Hubbard model.
Both the multiterminal conductance of an 
array of quantum dots and the persistent current of a quantum dot
molecule embedded in an Aharanov-Bohm ring are calculated.
The conductance and persistent current are 
calculated analytically for the case of a double quantum
dot and numerically for larger arrays
using a multi-terminal Breit-Wigner type formula, which allows for the 
explicit inclusion of inelastic processes.
Cotunneling corrections to the persistent current are also investigated, and
it is shown that the {\em sign} of the persistent current on resonance may
be used to determine the spin quantum numbers of the ground state and 
low-lying excitated states of an artificial molecule.
An inhomogeneous magnetic field is found to strongly suppress transport due 
to pinning of the spin-density-wave ground state of the system, and giant
magnetoresistance is predicted to result from the 
ferromagnetic transition induced by a uniform external magnetic field. 
\end{abstract}

\tighten
\widetext

\section{Introduction}

Interest in the problem of coherent resonant tunneling through an
interacting mesoscopic system has been stimulated by a series of elegant
Aharanov-Bohm (AB) ring experiments,\cite{yacoby} which measured the phase of
the transmission amplitude through a quantum dot in the Coulomb blockade 
regime.  Several theoretical works have addressed the role of phase coherence
in resonant tunneling through a single quantum dot:  both the conductance
\cite{cond.ab}
and the persistent current\cite{mbcs} of a quantum dot
embedded in an AB ring have been calculated.  Many features of the
experiments of Ref.\ \onlinecite{yacoby} have been explained by these
model calculations \cite{cond.ab,mbcs}; however, the correlations observed
between the phases of conductance resonances rather widely separated in
energy do not appear to be explicable within these simple models, and it is
therefore of interest to investigate coherent resonant tunneling through 
complex
mesoscopic systems with nontrivial substructure.  

In this article, we
investigate the multiterminal conductance of an artificial molecule
of coupled quantum dots as well as 
the persistent current of an artificial molecule embedded in an AB ring.
Arrays of coupled quantum dots 
\cite{dotarrays,dotsexp,vaart,waugh,pdots,blick,crouch} 
can be thought of as 
systems of artificial atoms separated by tunable tunnel barriers. 
The competition between intradot charge quantization effects,
or Coulomb blockade,\cite{onedot}
due to the ultrasmall capacitance of each quantum dot to its environment,
and coherent interdot
tunneling has been predicted to lead to a rich spectrum of many-body effects
in these systems.\cite{us,klimeck,twodot,us2,roza}  
For example, the Coulomb blockade of the individual quantum dots was 
predicted to be destroyed completely when interdot 
tunneling exceeds a critical value.\cite{us}  
This phenomenon can be considered a finite-size
analog of the Mott-Hubbard metal-insulator transition,\cite{us} and has
been observed experimentally in double quantum dots.\cite{waugh,crouch}
A detailed theoretical investigation of the double quantum dot in the limit
of a continuous energy spectrum on each dot, valid for very large dots, was 
able to describe the crossover from two isolated dots to a single, larger 
dot quite well.\cite{twodot}  An alternative approach using an effective
Hubbard-like model to describe the low-lying electronic states of the double
quantum dot, appropriate in the limit of a discrete energy spectrum on each dot,
was also able to reproduce the experimentally observed crossover, as well as the
nonlinear conductance of the system.\cite{roza}  Here we will employ such an
effective Hubbard model to describe coherent resonant tunneling through a
one-dimensional (1D) array of quantum dots coupled to multiple leads.
We will briefly consider the corresponding situation for the 
coherent resonant tunneling through a two-dimensional (2D) quantum dot 
array as well. 
We focus on the strongly-correlated regime, where interdot tunneling is too
weak to destroy the energy gap stemming from Coulomb blockade effects, and
where the energy spectrum on each dot is discrete.

An important consequence of coherent interdot tunneling in the 
strongly-correlated regime is the formation of
interdot spin-spin correlations \cite{pwa} 
analogous to those in a chemical bond
at an energy scale $J \sim  t^2/U$, where 
$U$ is the charging energy of a quantum dot  
and $t = (\hbar^2/2m^{\ast}) \int d^3 {\bf x} \,\Psi^{\ast}_m
({\bf x}) \nabla^2 \Psi_n ({\bf x})$ is the interdot hopping matrix element, 
$\Psi_{m,n}$ being electronic orbitals on nearest-neighbor dots.  
In a system with magnetic disorder, such a spin configuration is pinned,
and the resulting blockage of spin backflow \cite{thermo} leads to strong
charge localization.  However,
an applied magnetic field will break such an antiferromagnetic
bond when the Zeeman splitting $g\mu_B B > J$, leading to
an enormous enhancement of the charge mobility.
Such spin-dependent many-body 
effects on the magnetotransport should be experimentally
observable provided $\Gamma + \Gamma^{(i)} , k_B T \lesssim J$, where 
$\Gamma+\Gamma^{(i)}$ is the
total broadening of the resonant levels of the system due to finite lifetime
effects and inelastic scattering; 
they can be readily distinguished from orbital effects
in arrays of quasi-two-dimensional quantum dots
by applying the magnetic field {\em in the plane} of the dots.
Observation of the predicted giant magnetoresistance effect in the 
low-temperature transport through 
coupled quantum dots would, we believe, represent a clear signature of the
formation of an artificial molecular bond.

The paper is organized as follows:  In Sec.\ II, an extended Hubbard model
describing the low-lying electronic states of an array of coupled quantum dots
is introduced, and the magnetic phase diagram of the system is discussed.
In Sec.\ III, some general expressions for the conductance and persistent 
current of an interacting mesoscopic system coupled to multiple electron
reservoirs are derived.  In Sec.\ IV, the conductance through a double quantum
dot in an inhomogeneous magnetic field is calculated.  The magnetoresistance
of the system is shown to be proportional to $U^3/t^4$.
The persistent current through a
double quantum dot embedded in an AB ring is also investigated.  For
a system with an odd number of electrons, it is shown that resonant tunneling 
through molecular states with odd and even $S_z$ leads to contributions to the 
persistent current of opposite signs.  In Sec.\ V, the conductance of
one-dimensional arrays of quantum dots is investigated.  The inelastic
scattering rate in the system is shown to be suppressed in large arrays
due to the orthogonality catastrophe.  A pronounced suppression of certain
resonant conductance peaks in an applied magnetic field is predicted to 
result from a field-induced ferromagnetic transition.  A many-body
enhancement of localization is predicted to give rise to a giant
magnetoresistance effect in systems with spin-dependent disorder.
Such spin-dependent magnetoresistance effects are found to be much weaker
in the ballistic transport regime.
In Sec.\ VI the conductance in the 2D quantum dot array is investigated, 
and the qualitative physics is found to be similar to that in the 1D array. 
Some conclusions are given in Sec.\ VII.

\section{Model}

\begin{figure}
\vbox to 9.5cm {\vss\hbox to 17cm
  {\hss\
    {\includegraphics{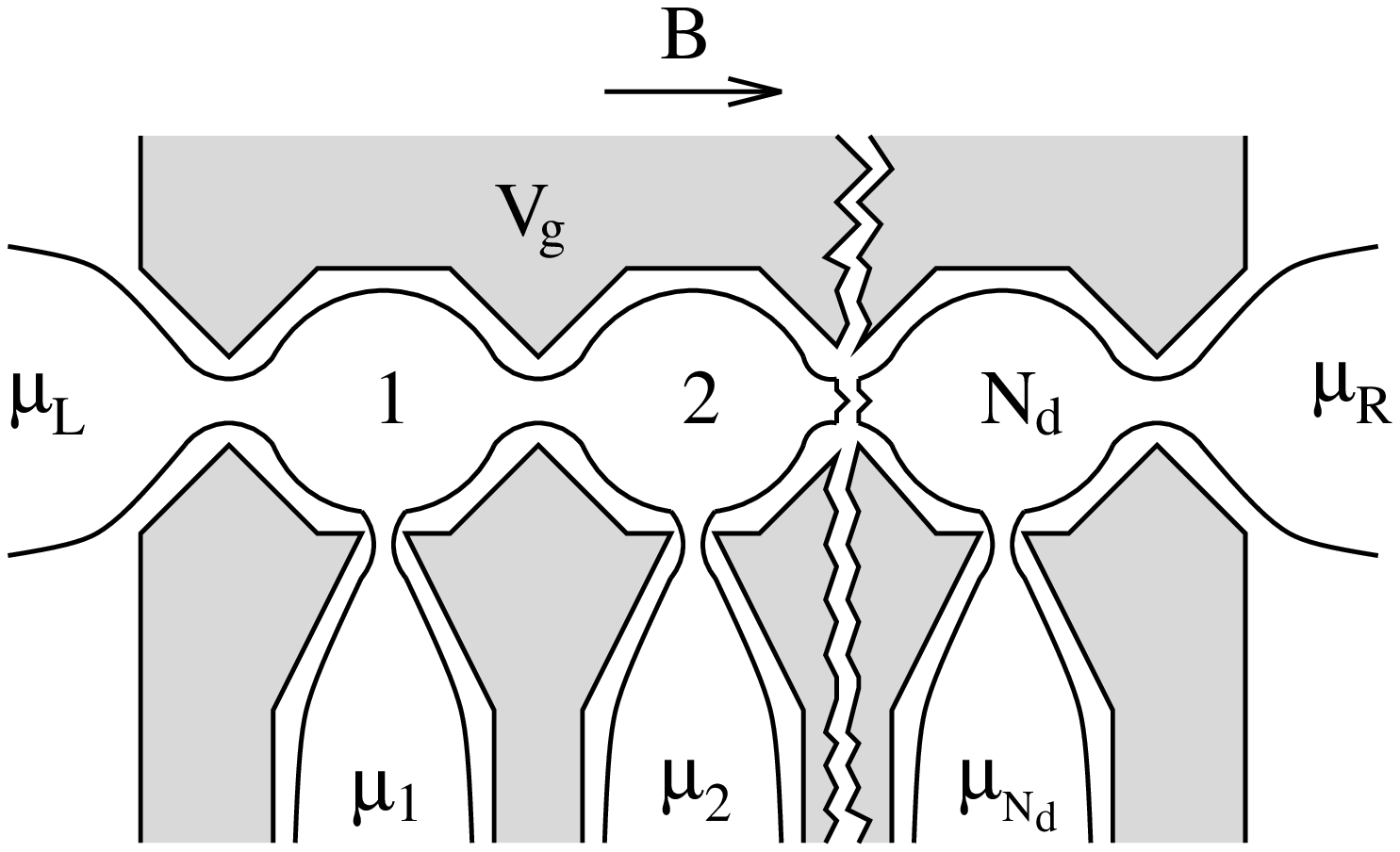}}
   \hss}
}
\vbox to 9.5cm {\vss\hbox to 17cm
  {\hss\
    {\includegraphics{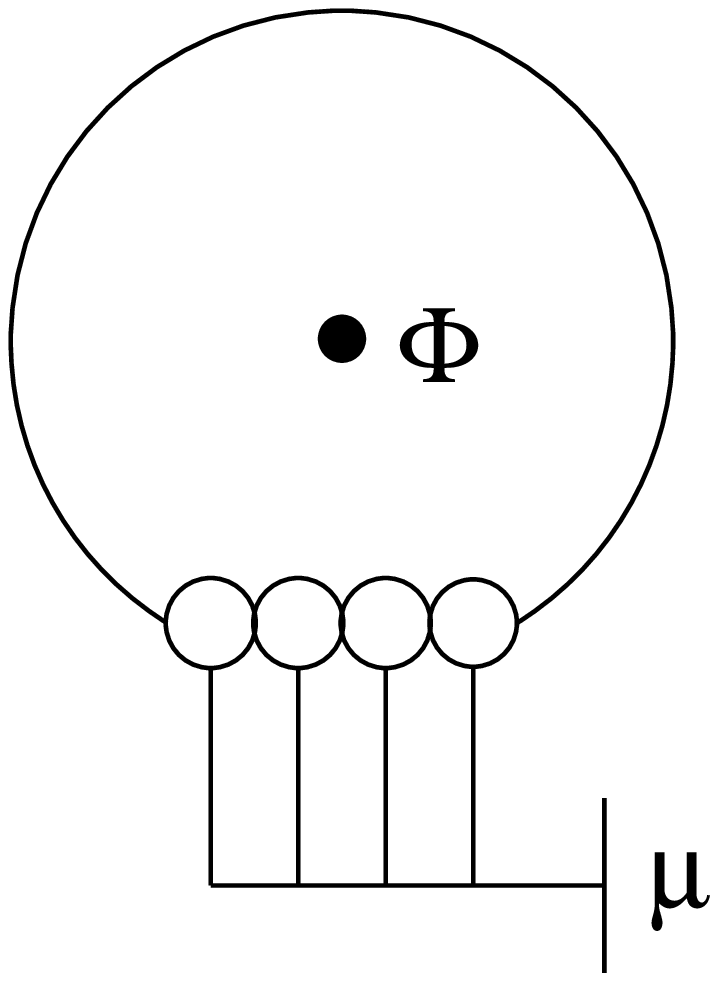}}
   \hss}
}
\caption{(a) Schematic diagram of a linear array of quantum dots.
(b) A quantum dot array embedded in an Aharanov-Bohm ring, formed by 
connecting the left and right reservoirs in (a).}
\label{system}
\end{figure}

The system under consideration (Fig.\ \ref{system})
consists of a linear array of quantum dots
electrostatically defined \cite{dotsexp,vaart,waugh,pdots,blick} 
in a 2D electron gas, 
coupled weakly to several macroscopic electron reservoirs, with a  
magnetic field in the plane of the dots.  
Each quantum dot is modeled by a single
spin-1/2 orbital, representing the electronic
state nearest the Fermi energy $E_F$, and is coupled via tunneling to its
neighbor and to one or more electron reservoirs.  
Transport occurs between the left (L) and right (R) reservoirs;
reservoirs 1 to $N_d$ are considered to
be ideal voltage probes,\cite{probe} and serve to 
introduce inelastic processes in the system.\cite{dephase}
Electron-electron interactions in the array are described 
\cite{onedot,classicaldots} 
by a matrix of capacitances $C_{ij}$:  We assume 
a capacitance $C_g$ between each quantum dot and the system of metallic
gates held at voltage $V_g$,  
an interdot capacitance $C$, and a capacitance $C_r$ between a quantum 
dot and each of its associated electron reservoirs.
The diagonal elements of $C_{ij}$
are the sum of all capacitances associated with a quantum dot,
$C_{11}=C_{N_d N_d}=C_g+C+2C_r$, $C_{ii}=C_g+2C+C_r$ ($1<i<N_d$) and the
off-diagonal elements are $C_{ij}=-C$ for nearest neighbor dots $ij$.
These capacitance coefficients may differ from their geometrical 
values due to quantum mechanical corrections,
\cite{twodot,cap1} but enter only as parameters in our model.
The Hamiltonian of the quantum dot array is
\begin{equation}
H_{\rm dots}  =  \sum_{j,\sigma}  \epsilon_{j\sigma}
d^{\dagger}_{j\sigma} d_{j\sigma}
+ \sum_{j,\sigma}
\left(t_{j\sigma} d^{\dagger}_{j+1\sigma} d_{j\sigma}
 + \mbox{H.c.}\right)
 + \frac{1}{2} \sum_{i,j} (Q_i + Q_g) C^{-1}_{ij} (Q_j + Q_g)
 - \frac{Q_g^2}{2} \sum_{i,j} C_{ij}^{-1},
\label{hubham}
\end{equation}
where $d^{\dagger}_{j \sigma}$ creates an
electron of spin $\sigma$ in the $j$th dot,
$Q_{j} \equiv -e\sum_{\sigma} 
d_{j \sigma}^{\dagger} d_{j \sigma}$ is the charge operator for dot $j$,
$Q_g \equiv C_g V_g$ is a polarization charge induced by the gate, and
\begin{equation}
\epsilon_{j \sigma}  = \epsilon_{j} + \sigma B_j/2 -e C_r\sum_i C_{ji}^{-1}
V_i, 
\label{site.energy}
\end{equation}
where $B_j$ is the Zeeman splitting on dot $j$  
and $\epsilon_{j}$ is the orbital energy of the quantum-confined orbital
under consideration on dot $j$.  The last term in Eq.\ (\ref{site.energy})
represents a shift in the orbital energy due to the capacitive coupling to
the reservoirs.  This term has an important effect on the nonlinear 
transport,\cite{nonlinear} but does not affect the linear response 
and equilibrium properties which are the subject of the present article.
We therefore set $C_r=0$ in the following.

$H_{\rm dots}$ reduces to
a Hubbard model \cite{us} with on-site repulsion $U=e^2/C_g$ in the limit
$C\rightarrow 0$.  In general, $H_{\rm dots}$
describes an extended Hubbard model with screened long-range interactions.
The elements of the inverse capacitance matrix 
decrease exponentially with a screening length that increases with 
$C/C_g$.  For $C\ll C_g$, $C_{ij}^{-1} \sim (C/C_g)^{|i-j|}/C_g$,
while for $C/C_g \rightarrow \infty$, $C_{ij}^{-1} \rightarrow 1/N_d C_g$,
and intradot charging effects are fully screened.  Electronic correlation 
effects are thus decreasing functions of $C/C_g$.   By varying the 
interdot electrostatic coupling, one can thus study the 
transition from a strongly correlated artificial molecule exhibiting 
collective Coulomb blockade\cite{us} for $C\ll C_g$ to a ballistic 
nanostructure where correlation effects are negligible for $C\gg C_g$.

\begin{figure}
\vbox to 17cm {\vss\hbox to 17cm
 {\hss\
   {\includegraphics{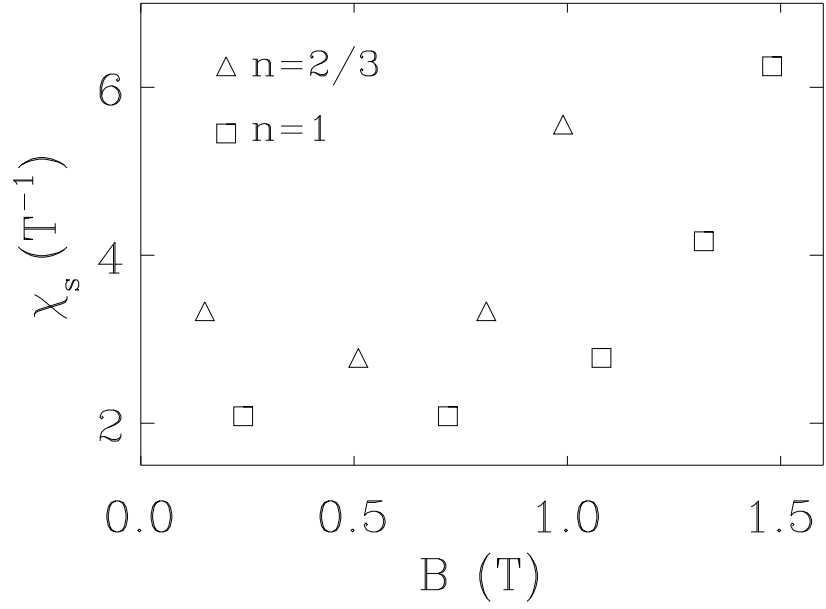} }
  \hss}
}
\caption{Spin susceptibility $\chi_s = \hbar^{-1} \Delta S/\Delta B$ at $T=0$
vs.\ magnetic field $B$ for linear arrays of GaAs
quantum dots with
$e^2/C_g=1\mbox{meV}$, $C/C_g=0.5$, and $t=.05\mbox{meV}$.  
Squares:  10 electrons on 10 dots ($B_c \approx 1.5 \mbox{T}$); 
triangles:  8 electrons on 12 dots  ($B_c \approx 1 \mbox{T}$).}
\label{chi_s}
\end{figure}

Of particular interest to us here is the magnetic phase diagram of the 
quantum dot array.  
In the strongly-correlated regime, where intradot charging is not strongly
screened, the $N$-electron ground state of $H_{\rm dots}$ will form a
spin-density-wave (SDW) due to interdot superexchange.\cite{pwa}
In an external magnetic field, the electron spins will tend to align with
the field to minimize the Zeeman energy.  There is thus a transition from
a spin-density-wave ground state at $B=0$ to a ferromagnetic state at some
critical magnetic field $B_c$.  For an infinite 1D array with $C=0$ and $t
\ll U$, one finds\cite{crit}
\begin{equation}
g\mu_{B} B_c \simeq \frac{4 t^2}{\pi U} (2 \pi n - \sin 2 \pi n),
\label{bcrit}
\end{equation}
where $n =1-|1-N/L|$ is the filling fraction of electrons (holes) for
$N<L$ ($N>L$) in the orbital under consideration.
Recall that
we are here considering only the single spin-1/2 orbital nearest $E_F$
in each quantum dot; the magnetic field required to spin-polarize an entire
quantum dot is much larger.\cite{klein}  
In a system with  $C > 0$, intradot Coulomb interactions are screened, and 
$B_c$ is therefore expected to increase.
Fig.\ \ref{chi_s} shows the spin susceptibility $\chi_s$
for $C/C_g=1/2$ in linear 
arrays with 8 electrons on 12 dots and 10 electrons
on 10 dots.  The $n$-dependence of $B_c$ in Fig.\ \ref{chi_s}
is qualitatively similar to that in a system with intradot interactions only,
but the values of $B_c$ are roughly twice
those of a system with $C=0$.
Note the rapid growth of $\chi_s$ as $B \rightarrow B_c$.
In an infinite array, $\chi_s$ is expected to diverge as $B \rightarrow B_c$
because the system undergoes a second order quantum phase 
transition.\cite{crit}  The spin-polarization transition (SPT)
predicted to occur in an array of coupled
quantum dots is in contrast to that observed in a single quantum dot,
\cite{klein} where the critical point occurs for minimum total spin.  

In the following, we shall investigate the effects of the SDW correlations 
and the SPT on the low-temperature magnetotransport through coupled
quantum dots.

\section{Quantum Transport Formalism}

Before investigating the particular Hamiltonian of interest, Eq.\ 
(\ref{hubham}), it will first be useful to derive some general formulas for the 
conductance and persistent current arising due to resonant tunneling
through an interacting system.  An arbitrary interacting
mesoscopic conductor coupled to $M$ macroscopic electron
reservoirs is described by the Hamiltonian
\begin{equation}
H = H_{\rm int}(\{d_n^{\dagger},d_n\}) +
\sum_{\alpha=1}^{M} \sum_{k \in \alpha}
\epsilon_k c_k^{\dagger} c_k
+ \sum_{\alpha=1}^{M}
\sum_{k\in\alpha}\sum_n
\left(V_{kn} c^{\dagger}_k d_n + {\rm H.c.}\right),
\label{ham}
\end{equation}
where $\{d^{\dagger}_n\}$ creates a complete set of single-particle 
states in the mesoscopic system, $c^{\dagger}_{k\in\alpha}$ 
creates an electron in state $k$ of reservoir $\alpha$, and $H_{\rm int}$
is a polynomial in $\{d^{\dagger}_n,d_n\}$ which commutes with 
the electron number $N=\sum_n d^{\dagger}_n d_n$.  Here the spin index 
$\sigma$ has been absorbed into the subscripts $n$ and $k$.
We denote the ground state of $H_{\rm int}$
for each $N$ by $|0_N\rangle$ and the ground state energy by $E_N^0$.  We
assume $E_N^0$ to be nondegenerate, as is generically the
case in a nonzero magnetic field.

If the tunneling barriers to the reservoirs are sufficiently large, and if
the temperature and bias
are small compared to the energy of an excitation, then
the main effect of particle exchange with 
the reservoirs will be to cause transitions
$|0_{N-1}\rangle \rightarrow |0_N\rangle$
between the nondegenerate ground states of the system.
In the vicinity of such a resonance,
the nonequilibrium Green's functions describing propagation within the
system in the presence of coupling to the leads can be shown to have the 
Breit-Wigner form\cite{nonlinear}
\begin{eqnarray}
G^r_{nm}(\epsilon) & = & \frac{\langle 0_{N-1}|d_{n}
|0_N\rangle
\langle 0_N|d^{\dagger}_{m}
|0_{N-1}\rangle}{\epsilon - E_N^0 + E_{N-1}^0 + i\Gamma_N/2}
+ \mbox{additional poles},
\label{gret}\\
G^{<}_{nm}(\epsilon) & = & \frac{i
\langle 0_{N-1}|d_{n} |0_N\rangle \langle 0_N|d^{\dagger}_{m} |0_{N-1}\rangle
\sum_{\alpha}\Gamma_N^{\alpha}f_{\alpha}(\epsilon)
}{(\epsilon - E_N^0 + E_{N-1}^0)^2 + (\Gamma_N/2)^2}
+ \mbox{additional poles},
\label{gless}
\end{eqnarray}
where $f_{\alpha}(\epsilon)=\{\exp[(\epsilon-\mu_{\alpha})/k_B T]+1\}^{-1}$ 
is the Fermi function for reservoir $\alpha$,
$\Gamma_N = \sum_{\alpha=1}^{M} \Gamma_N^{\alpha}$, and
\begin{equation}
\Gamma_N^{\alpha} = 2\pi \sum_{k\in\alpha}\sum_{n,m}
\langle 0_{N-1}|V_{kn}d_n|0_N\rangle
\langle 0_N |V^{\ast}_{km}d^{\dagger}_m |0_{N-1}\rangle
\delta(\epsilon_k - E_N^0 + E_{N-1}^0).
\label{gam}
\end{equation}
Here, $G^{<,r}_{nm}(\epsilon)$ 
are Fourier transforms of the Keldysh Green's function
$G^<_{nm}(t)=i \langle d^{\dagger}_{m}(0)
d_{n}(t)\rangle$ and the retarded Green's function
$G^r_{nm}(t)=-i\theta(t)
\langle \{d_{n}(t),d^{\dagger}_{m}(0)\}\rangle$, respectively.
With the aid of these Green's functions, the conductance and persistent
current resulting from resonant tunneling through the system can be calculated.

\subsection{Multiterminal Conductance Formula}

The expectation value of the current flowing out of the interacting
region into reservoir $\alpha$ 
can be expressed using the formalism of Meir and Wingreen as \cite{mw}
\begin{equation}
I_{\alpha} = -\frac{e}{h} \int 
d\epsilon\,\mbox{Im}\,\mbox{Tr}\left\{\Gamma^{\alpha}(\epsilon)
\left[G^{<}(\epsilon) +
2 f_{\alpha}(\epsilon) G^r(\epsilon)\right]\right\},
\label{current}
\end{equation}
where $\Gamma^{\alpha}_{nm}(\epsilon)=2\pi\sum_{k\in\alpha}
V_{kn}V^{\ast}_{km} \delta(\epsilon-\epsilon_k)$ is a matrix characterizing
the tunnel barrier connecting reservoir $\alpha$ to the system.
Inserting $G^{<,r}_{nm}$ from Eqs.\ (\ref{gret}) and (\ref{gless})
into Eq.\ (\ref{current}), one finds the multiprobe current formula
for resonant tunneling\cite{nonlinear}
\begin{equation}
I_{\alpha}
= \frac{e}{h} \sum_{\beta=1}^{M}
\int d\epsilon \sum_N \frac{
\Gamma_N^{\alpha}\Gamma_N^{\beta}\, 
[f_{\alpha}(\epsilon)-f_{\beta}(\epsilon)]
}{\left(\epsilon - E_N^0 + E_{N-1}^0\right)^2
+ \left(\Gamma_N/2\right)^2}.
\label{multi}
\end{equation}
The low-temperature transport through such a correlated many-body system
weakly coupled to multiple leads
thus exhibits resonances of the Breit-Wigner type,\cite{dephase}
where the positions and intrinsic
widths of the resonances are determined by the {\em many-body} 
states of the system.  
Eq.\ (\ref{multi}), which expresses the current in terms of transmission
probabilities, is a generalization of the multi-terminal conductance
formula for a noninteracting system derived by B\"uttiker \cite{mb} 
to the case of resonant tunneling through an interacting system.  

In deriving Eq.\ (\ref{multi}), we have neglected the
additional poles in $G^{<,r}_{nm}(\epsilon)$, 
which is justified provided $k_B T,\,
\Gamma_N \ll \Delta E_N$ and $\Delta \mu < \Delta E_N$, where
$\Delta E_N=\min(E^1_N-E^0_N,E^1_{N-1}-E^0_{N-1},E^0_{N+1}-E^0_N-\mu_{\alpha},
\mu_{\alpha}-E^0_{N-1}+E^0_{N-2})$,
$E^1_N$ being the energy of the lowest lying excited
state of the $N$-electron system.
Eq.\ (\ref{multi}) is thus appropriate to describe 
resonant tunneling through semiconductor nanostructures \cite{dotexp1,dotexp2}
or ultrasmall metallic/superconducting systems \cite{scdot,noandreev} 
under conditions of
low temperature and bias, where transport is dominated by a {\em
single} ground-state to ground-state transition $|0_{N-1}\rangle
\leftrightarrow |0_N\rangle$.
Eq.\ (\ref{multi}) is not applicable
to systems with a (spin)degenerate ground state [$\Delta E(N)=0$], where
the low-temperature physics is that of the Kondo effect, as discussed in
Refs.\ \onlinecite{mw,kondot,bfs,assh}.

We next specialize to the configuration shown in 
Fig.\ \ref{system}.
Transport occurs between the left ($L$) and right ($R$) reservoirs. 
The auxiliary reservoirs $1,\ldots,N_d$ are assumed to be connected to
ideal voltmeters.\cite{probe}
An ideal voltmeter should have an infinite impedance,
so we demand that the expectation value of the current 
flowing into reservoirs $1,\ldots,N_d$ be zero, which fixes 
$\mu_1,\ldots,\mu_{N_d}$ via Eq.\ (\ref{multi}).
Eliminating $f_1(\epsilon),\ldots, f_{N_d}(\epsilon)$ from Eq.\ (\ref{multi}),
and taking the linear response limit, one finds the effective two-terminal
conductance between the left and right  contacts
\begin{equation}
G = \frac{e^2}{h} \sum_N \frac{\Gamma_N^L \Gamma_N^R}{\Gamma_N^L + \Gamma_N^R}
\int \frac{ \Gamma_N\, [-f'(\epsilon)]\, d\epsilon 
}{\left(\epsilon - E_N^0 + E_{N-1}^0\right)^2
+ \left(\Gamma_N/2\right)^2}.
\label{conductance}
\end{equation}
The total width of the $N$th resonance may be written
$\Gamma_N = \Gamma_N^L + \Gamma_N^R
+ \Gamma_N^{(i)}$, where 
$\Gamma_N^{(i)} = \sum_{\alpha=1}^{N_d} \Gamma_N^{\alpha}$.
The quantity $\Gamma_N^{(i)}/\hbar$ may be interpreted 
as the total {\em inelastic scattering rate} due to 
phase-breaking processes in the auxiliary reservoirs.\cite{dephase}
Such processes arise when an electron in the $i$th dot 
escapes into reservoir $i$ and is replaced by another electron from the
reservoir, whose phase is uncorrelated with that of 
the previous electron.\cite{dephase}

\begin{figure}
\vbox to 17cm {\vss\hbox to 17cm
  {\hss\
    {\includegraphics{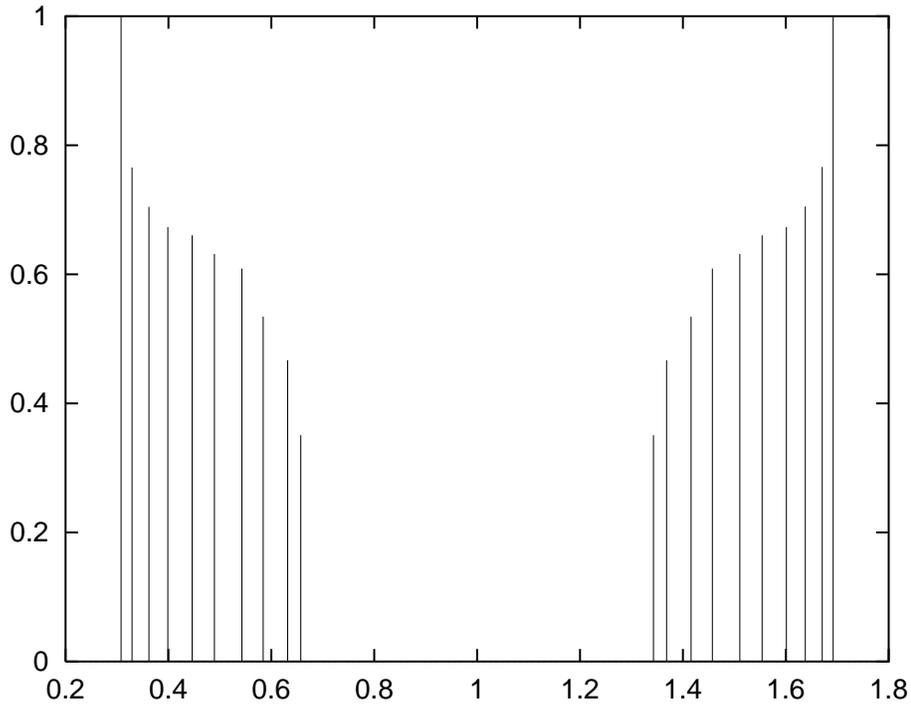}}
   \hss}
}
\caption{The total wavefunction overlap $\Gamma_N^{(i)}/\Gamma^{(i)} =
\sum_{i=1}^{N_d} \sum_{\sigma} |\langle 0_N|d^{\dagger}_{i\sigma}|0_{N-1}
\rangle|^2$ as a function of $N$ for an array of ten quantum dots with
pure Hubbard interactions $U=10t$.  The $x$ coordinate of the $N$th
peak indicates the value of the polarization charge $Q_g=C_g V_g$ at which 
the $N$th electron is added to the system.  The suppression of 
$\Gamma_N^{(i)}$  with increasing $N$ is evident in the lower Hubbard band
$Q_g/e < 1$.  The symmetry about $Q_g/e=1$ follows from the particle-hole
symmetry of Eq.\ (\ref{hubham})}
\label{ortho.fig}
\end{figure}

For simplicity, let us assume that the tunnel barriers 
coupling the system to the external reservoirs are described by the
energy-independent parameters
$2\pi\sum_{k\in\alpha}|V_{kn}|^2 \delta(\epsilon_k-E) =
\Gamma^{(i)}\delta_{n\alpha}$, $\alpha=1,\ldots,N_d$;  
$\Gamma\delta_{n1}$, $\alpha=L$; 
$\Gamma\delta_{nN_d}$, $\alpha=R$. 
Then the partial widths of the $N$th resonance are simply 
$\Gamma^{L}_N = \Gamma \sum_{\sigma} |\langle 0_N|d^{\dagger}_{1 \sigma}
|0_{N-1}\rangle|^2$, 
$\Gamma^{R}_N = \Gamma \sum_{\sigma} |\langle 0_N|d^{\dagger}_{N_d \sigma}
|0_{N-1}\rangle|^2$, and
\begin{equation}
\Gamma_N^{(i)} = \Gamma^{(i)} \sum_{i=1}^{N_d} \sum_{\sigma} |\langle 0_N|
d^{\dagger}_{i\sigma}|0_{N-1}\rangle|^2.
\label{gamm.i}
\end{equation}
Since $\{d^{\dagger}_{i\sigma}\}$ creates a complete basis of single-particle
states in the array, one would have $\Gamma_N^{(i)} = \Gamma^{(i)}$ for a
noninteracting system.  However, in an interacting system the 
wavefunction overlap $|\langle 0_N|d^{\dagger}_{i\sigma}|0_{N-1}\rangle|$
is suppressed by correlation effects, leading to an orthogonality catastrophe
in the large-$N$ limit.\cite{ortho}  The many-body 
suppression of the wavefunction overlaps leads to a reduction of both the 
elastic broadening $\Gamma_N^{(e)}=\Gamma_N^L + \Gamma_N^R$ 
and of the inelastic broadening
$\Gamma_N^{(i)}$ of the conductance resonances, so that the system becomes
more and more weakly coupled to the environment as $N$ increases.
The suppression of the wavefunction overlap with increasing $N$ is shown for
an array of ten quantum dots with pure Hubbard interactions in Fig.\ 
\ref{ortho.fig}.
The effect of such an orthogonality catastrophe in the sequential tunneling
regime has previously been discussed by
Kinaret {\it et al.} \cite{ortho2} and by Matveev, Glazman, and Baranger.
\cite{twodot}

\subsection{Persistent current}

Let us next consider the persistent current through the quantum dot array
when the left and right reservoirs in Fig.\ 1(a) are connected together to form
a 1D ring of length $L$ enclosing an AB flux $\Phi=(\hbar c/e)\phi$.  
Since the persistent current is an equilibrium property, the auxiliary
reservoirs must be in mutual equilibrium at electrochemical potential $\mu$,
as indicated in Fig.\ 1(b).
The persistent current in such an open system is given by 
$I(\phi)= -(e/\hbar) \partial \Omega/\partial\phi$,
where $\Omega$ is the grand canonical potential.   The grand canonical
potential may be determined from the electronic scattering matrix of the 
system,
as discussed by Dashen, Ma, and Bernstein \cite{dashen} and by
Akkermans {\it et al.},\cite{akkermans}  and is
\begin{equation}
\Omega(\phi)= -\int \frac{dE}{2\pi} f(E) \, \mbox{Im} \ln \det S(E,\phi),
\label{omega}
\end{equation}
where $S(E,\phi)$ is the scattering matrix of the multiply-connected structure
shown in Fig.\ 1(b) and $f(E)$ is the Fermi function with electrochemical
potential $\mu$.  
In order to obtain $S(E,\phi)$, it is useful first to 
introduce the scattering matrix $S(E)$ for the structure shown in Fig.\ 1(a),
which is related simply to the retarded Green's function,
\begin{equation}
S_{kk'}(E) = -\delta_{kk'} + 2\pi i \sum_{nm} \frac{V_{k m}^{\ast}
V_{k' n}}{\sqrt{v_{k} v_{k'}}} G^r_{nm}(E),
\label{s.dots}
\end{equation}
where $v_{k\in\alpha}$ is the group velocity of state $k$ in reservoir 
$\alpha$.  Using Eq.\ (\ref{gret}), one sees that $S(E)$ has the Breit-Wigner
form discussed by B\"uttiker.\cite{dephase}  $S(E)$ may be divided into
submatrices $r_{kk'}$, describing reflection of mode $k'$ in the reservoir
back into mode $k$ in the reservoir, $\varepsilon_k$ and $\delta_k$,
describing transmission of mode $k$ from the reservoir 
into the the ring to the left and right, respectively, 
$\gamma$, describing transmission of a circulating
state in the ring through the quantum dot array, and $\alpha$ and $\beta$,
describing reflection of the circulating states in the ring at the left and
right ends of the quantum dot array, respectively.   In terms of the
submatrices of $S(E)$, the flux-dependent
scattering matrix for the combined structure may be written
\begin{equation}
S_{kk'}(E,\phi) = r_{kk'} + \frac{\alpha \delta_k \delta_{k'}
+\beta\varepsilon_k \varepsilon_{k'} - (\gamma-\cos\phi e^{-ipL})
(\varepsilon_k \delta_{k'}+\delta_k\varepsilon_{k'})}{
\gamma^2-\alpha\beta-2\gamma\cos\phi e^{-ipL} + e^{-2ipL}},
\label{sofphi}
\end{equation}
where $p=\sqrt{2mE}/\hbar$ and it is assumed that $\epsilon_k=\epsilon_{k'}
=E$.   Substituting Eq.\ (\ref{sofphi}) into Eq.\ (\ref{omega}), and taking 
the derivative with respect to $\phi$, one obtains the persistent current
due to coherent resonant tunneling through the quantum dot array.
The general expression for $I(\phi)$ is somewhat cumbersome, but for the
special case $\Gamma_N^L=\Gamma_N^R=\Gamma_N^{(e)}/2$, one finds
the simple result
\begin{equation}
I(\phi)=\frac{e\Gamma_N^{(e)}}{2h}
\int \frac{dE\,f(E)\, \Gamma_N^{(i)} \frac{\textstyle
\sin\phi\sin pL}{\textstyle (\cos\phi -
\cos pL)^2}}{\left[E-E_N^0+E_{N-1}^0-\frac{(\Gamma_N^{(e)}/2)\sin pL}{
\cos pL -\cos\phi}\right]^2 + \left(\Gamma_N^{(i)}/2\right)^2}.
\label{i.huge}
\end{equation}
The persistent current is thus also determined by the conductance matrix
elements $\Gamma_N^{\alpha}$.  From Eq.\ (\ref{i.huge}) one sees that as 
the energy of the resonance is tuned through the Fermi level (by varying
the gate voltage $V_g$), the persistent current exhibits a maximum of
height $\mbox{} \sim (e\Gamma_N^{(e)}/\hbar)\sin \phi \sin k_F L$, which
decays with a width $\Gamma_N$ when the resonance is moved above $\mu$.
$I$ also decays when the resonance is moved below the Fermi level, but in an
oscillatory pattern determined by the level spacing in the ring.

In the limit $\Gamma_N^{(i)}\rightarrow 0$ (closed system), 
the integrand in Eq.\ (\ref{i.huge}) consists of a series of delta functions 
whose weights give the current contributed by each of the discrete states in
the ring which couple to the dot array, subject to periodic boundary 
conditions.
If the ring has dimensions comparable to those of the dot array, then the 
level spacing in the ring may exceed the width of the resonance ($\hbar v_F/L
> \Gamma_N$), and the above approach will break down.  The 
persistent current will then be dominated by charge fluctuations coupling the
highest occupied state in the ring, of energy $\epsilon_F$,
with the lowest unoccupied many-body 
state in the dotarray.\cite{mbcs}  The coupling matrix element is
\begin{equation}
|t|^2 = t_R^2 + t_L^2 + 2 t_R t_L \cos \phi,
\label{t.pm}
\end{equation}
where the matrix elements
\begin{eqnarray}
t_L & = & \sum_{\sigma} 
\langle 0_N| V_{k_F 1} d^{\dagger}_{1\sigma}|0_{N-1}\rangle
\label{t.l}\\
t_R & = & \sum_{\sigma} 
\langle 0_N| V_{k_F N_d} d^{\dagger}_{N_d\sigma}|0_{N-1}\rangle
\label{t.r}
\end{eqnarray}
may be chosen real.   A straightforward calculation\cite{mbcs} then yields
the persistent current through the quantum dot array
\begin{equation}
I(\phi) = -\frac{2(e/\hbar)t_R t_L \sin \phi}{[(\epsilon_F -
E_N^0+E_{N-1}^0)^2 + 4|t|^2]^{1/2}}.
\label{i.res}
\end{equation}
The product $t_R t_L$ may be greater or less than
zero depending on the relative signs of $V_{k_F 1}$ and $V_{k_F N_d}$, and on
the relative signs of the wavefunction overlaps, so the persistent current
on resonance may be either diamagnetic or paramagnetic.  For a purely 1D
system of spinless electrons, or with an even number of spin-1/2 electrons,
the sign of $I(\phi)$ is determined by the total number of electrons in the
system, as discussed in Ref.\ \onlinecite{mbcs}.  This is a manifestation
of the so-called Leggett theorem.\cite{leggett}
However, in general the sign of $I(\phi)$ depends on the particular state which
dominates the resonance, and may be used to classify the quantum numbers of
that state, as discussed below.

\section{Double Quantum Dot}

\begin{figure}
\vbox to 17cm {\vss\hbox to 17cm
  {\hss\
    {\includegraphics{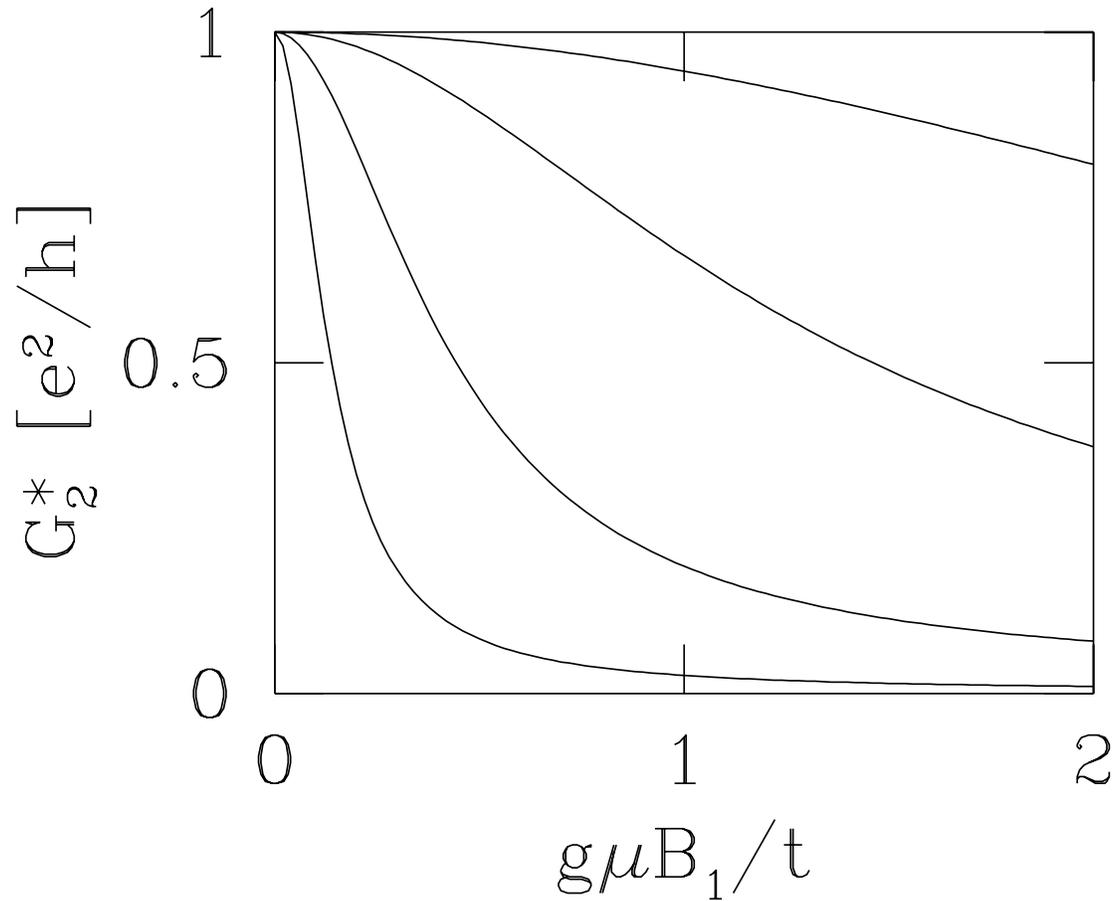}}
   \hss}
}
\caption{Resonant conductance at $T=0$ 
in units of $(e^2/h)/(1+\Gamma^{(i)}/
\Gamma)$ as a function of the Zeeman splitting on dot 1 for $\gamma^{-1}=0$, 3,
10, 30 (top to bottom).  If an additional uniform external field 
$B > J$ is applied, the conductance is restored to the 
value for $\gamma^{-1}=0$, leading to giant magnetoresistance.}
\label{GmaxofB}
\end{figure}

Let us first consider the case of double quantum dot, for
which the conductance matrix elements (\ref{gam}) can be obtained 
analytically.  For $N_d=2$, $H_{\rm dots}$ reduces to a two-site Hubbard
model with on-site repulsion $U=e^2 C_{\Sigma}/(C_{\Sigma}^2-C^2)$ and 
nearest neighbor repulsion $U_{12}= e^2 C/(C_{\Sigma}^2-C^2)$, where
$C_{\Sigma}=C_g + C + 2C_r$.
Spin disorder is introduced via a Zeeman splitting on dot 1,
$B_1=4t\Delta$, $B_2=0$. Experimentally, such an inhomogeneous field could
be produced, e.g., by the presence of a small ferromagnetic particle.
The total width of the one-particle resonance is found to be
$\Gamma_1=\Gamma+\Gamma^{(i)}$, and the 
prefactor in Eq.\ (\ref{conductance}) is
$\Gamma_1^L \Gamma_1^R/(\Gamma_1^L + \Gamma_1^R) = (\Gamma/4)/(
1+\Delta^2)\equiv \Gamma_0$.  The maximum conductance at the
one-particle resonance is thus
\begin{eqnarray}
G_1^* & = &
\frac{\textstyle e^2/h}{\textstyle (1+\Delta^2)(1+\Gamma^{(i)}/\Gamma)},
\;\;\;\;\;\;\; T=0 \nonumber\\
& = &
e^2 \Gamma_0/4\hbar k_B T, \;\;\;
\Gamma + \Gamma^{(i)} \ll k_B T \ll t\sqrt{1+\Delta^2}. 
\label{G1max}
\end{eqnarray}
Inelastic scattering suppresses the resonant conductance at $T=0$, but has
no effect when the resonance is thermally broadened.  For $U-U_{12}\gg t$,
the two-particle ground state of the double quantum dot
has an antiferromagnetic spin configuration
characterized by the superexchange parameter
\begin{equation}
J=2t\left(\gamma-\Delta+\sqrt{\gamma^2+\Delta^2}\right),
\label{exchange}
\end{equation}
where $\gamma=t/(U-U_{12})$.  Note that $2t\gamma \leq J \leq 4t\gamma$.
The two-particle resonance is separated from the one-particle resonance by
$e\Delta Q_g/(C_{\Sigma}-C) = U_{12} + 2t(1+\Delta^2)^{1/2} - J$,
and the conductance is determined by the matrix elements
\begin{equation}
\langle 0_2 | d^{\dagger}_{j\uparrow} |0_1\rangle =
\frac{\sqrt{2}}{A} \left\{\frac{2\gamma}{\left[1+\left(\Delta \mp \sqrt{
\Delta^2+1}\right)^2\right]^{1/2}} + \frac{1\mp \Delta/
\left(\gamma+\sqrt{\gamma^2 +
\Delta^2}\right)}{\left[1+\left(\Delta\pm\sqrt{\Delta^2+1}
\right)^2\right]^{1/2}}\right\},
\label{injectamp}
\end{equation}
where $A^2=1+\Delta^2/(\gamma+\sqrt{\gamma^2+\Delta^2})^2$ and 
the upper (lower) sign holds for $j=1$ (2).
For $B_1\gtrsim J$, the antiferromagnetic
spin configuration is pinned, leading to a strong
suppression of the amplitude to inject electron 2 into dot 1, and 
a concommitant suppression of the second conductance peak.  
Inserting Eq.\ (\ref{injectamp}) into
Eqs.\ (\ref{gam}) and (\ref{conductance}), one finds the $T=0$ resonant 
conductance
\begin{eqnarray}
G_2^* & = &
 16(e^2/h)(\gamma/\Delta)^2/(1+\Gamma^{(i)}/\Gamma),
\;\;\;\;\;\;\gamma \ll \Delta \ll 1 \nonumber \\
& = &
4(e^2/h)\gamma^2/(1+\Gamma^{(i)}/\Gamma),\;\;\;\;\;\; \Delta \gg 1.
\label{G2max}
\end{eqnarray}
A second doublet of conductance peaks
for $N=3,\,4$ is separated from this doublet by $\Delta Q_g\simeq e$ (center to
center), and one
finds $G_3^*=G_2^*$, $G_4^*=G_1^*$ due to electron-hole symmetry.  
The resonant conductance for $N=2$ is suppressed by a factor of
$\gamma^2$ compared to that for $N=1$ due to collective spin pinning
(one readily verifies that the resonant conductance is suppressed by the same 
many-body factor in the regime of thermally broadened resonances).
This dramatic many-body suppression
of the conductance is illustrated in Fig.\ \ref{GmaxofB} for several values 
of $\gamma$.
The effect of spin disorder is to be contrasted
with that of a charge detuning $\Delta=(\epsilon_1-\epsilon_2)/2t$, 
investigated by Klimeck {\it et al.} \cite{klimeck} and by
van der Vaart {\it et al.},\cite{vaart} for
which both $G_1^*$ and $G_2^*$ are given by Eq.\ (\ref{G1max}) at $T=0$
($G_2^*$ is then reduced by a factor of 2 in the thermally broadened regime).
The very different effects of spin and charge disorder stem from the fact
that the repulsive interactions in Eq.\ (\ref{hubham}) enhance spin-density
fluctuations, but suppress charge-density fluctuations.  

Let us now
consider the effect of an additional homogeneous magnetic field applied
parallel to the inhomogeneous field, $B_1=4t\Delta+B$, $B_2=B$.
For $B>J$, it is energetically favorable to break the antiferromagnetic
bond between the dots and form a spin-polarized state, thus
preventing collective spin pinning effects. $G_2^*$ is then
given by Eq.\ (\ref{G1max}). The resulting magnetoresistance on resonance for
$T=0$ and $\Delta\gg 1$ is thus
\begin{equation}
\frac{\Delta R^*}{\Delta B}= 
- \frac{h}{e^2}\frac{g\mu_B}{4J\gamma^2} (1+\Gamma^{(i)}/\Gamma)
\sim - \frac{h}{e^2}\frac{g\mu_B (U-U_{12})^3}{8t^4}.
\label{dRdB}
\end{equation}
In the thermally broadened resonance regime, the factor $(1+\Gamma^{(i)}/
\Gamma)$ is replaced by $2k_B T/\pi\Gamma$.
Since the Coulomb energy $U-U_{12}$ is typically large compared to the
interdot tunneling matrix element $t$, the predicted magnetoresistance
is extremely large.
This giant magnetoresistance effect is a direct indication of the
field-induced breaking of the artificial molecular bond between the dots.
\cite{parallel}

The conditions necessary to observe the predicted magnetoresistance effect 
may be determined by including the effect of transport through the triplet
excited state via the method of Refs.\ \onlinecite{onedot} and
\onlinecite{klimeck}.  One finds 
the resonant conductance at $B=0$ for $k_B T \gg \Gamma+\Gamma^{(i)}$
\begin{equation}
G_2^* = \frac{e^2}{2\hbar k_B T}\frac{\exp(\beta J)}{2\exp(\beta J)-1}
\left(\Gamma_s + \frac{2\Gamma_t}{\exp(\beta J)+1}\right),
\label{GofT}
\end{equation}
where $\Gamma_s\simeq 4\gamma^2\Gamma$ is the sequential tunneling
rate through the pinned antiferromagnetic ground state and $\Gamma_t
=\Gamma_0$ is the sequential tunneling rate through the triplet
excited state.  The magnetoresistance is thus reduced by a factor of 2 at a 
temperature $k_B T_{1/2} = J/\ln(\Gamma_t/\Gamma_s)$.
Increased coupling to the leads and/or inelastic scattering
can be shown 
to lead to a similar admixture of transport through excited
states when $\Gamma+\Gamma^{(i)}\sim J$.
We therefore expect the predicted giant magnetoresistance
effect to be observable for $k_B T,\,\Gamma +\Gamma^{(i)}\lesssim J$.  
In currently
available GaAs quantum dot systems, charging energies are typically of
order 1meV, and one expects tunneling matrix elements $t\sim 0.1\mbox{meV}$ 
for moderate to strong interdot tunneling, so 
values of $J$ in the range .01---0.1meV should be attainable.

\begin{figure}
\vbox to 17cm {\vss\hbox to 17cm
  {\hss\
    {\includegraphics{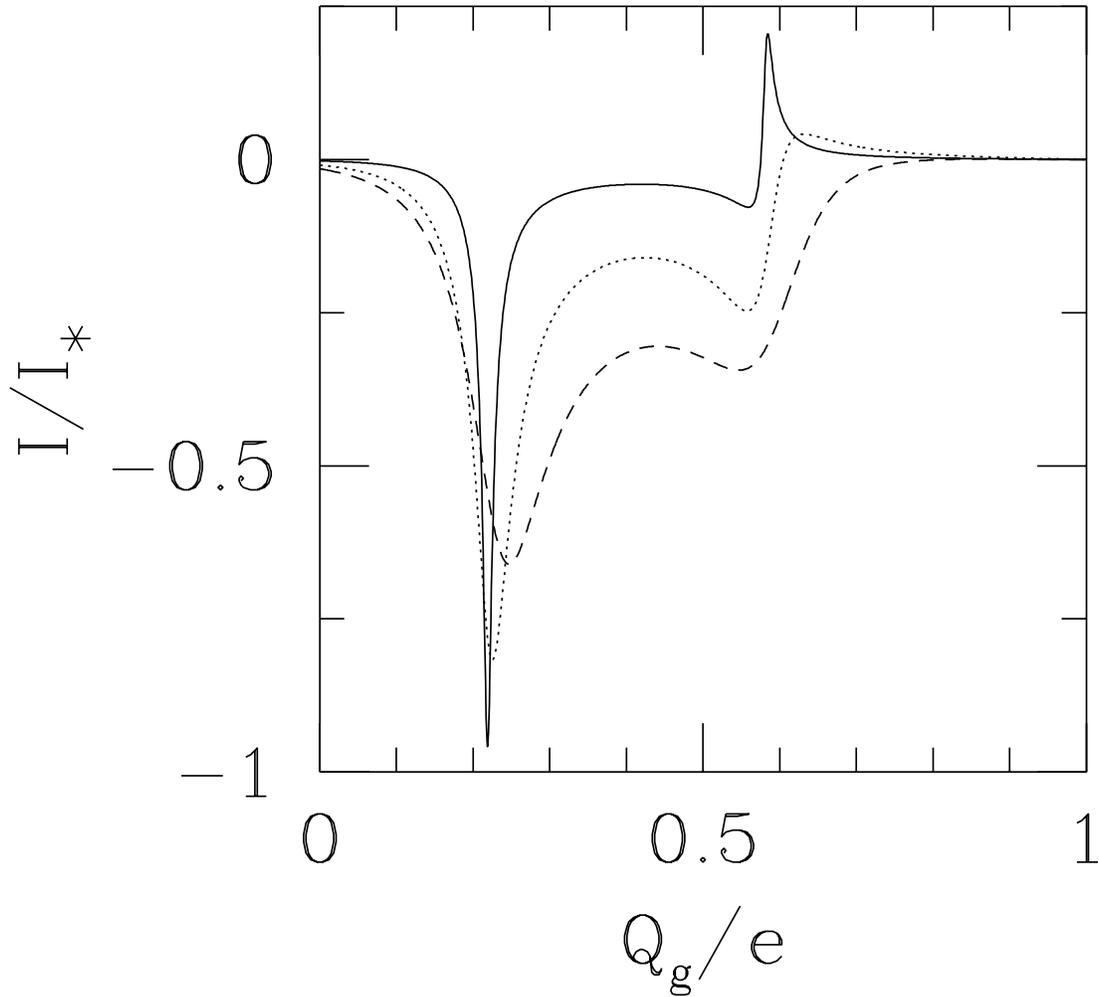}}
   \hss}
}
\caption{Persistent current through a double quantum dot embedded in an
Aharanov-Bohm ring as a function of the gate voltage.  Here $t=1$,
$U=10$, $U_{12}=0$, and $V_{k1}=V_{kN_d}=V$.  The current is expressed in 
units of $I_{\ast}=e V/2^{1/2} \hbar$, the value on resonance in the absence
of correlations and asymmetry.
Solid curve:  $V=1/32$; dotted curve: $V=1/8$; dashed curve: $V=1/2$.
}
\label{ires.lanczos}
\end{figure}

Let us next consider the persistent current through the double quantum
dot.  From Eqs.\ (\ref{i.huge}) and (\ref{i.res}), one sees that the 
persistent current is also suppressed at the $N=2$ resonance due to the 
many-body factor (\ref{injectamp}).  However, in the later case of a 
nanoscopic ring with level spacing $\hbar v_F/L \gg \Gamma_2$, the suppression
of the persistent current is only linear in $\langle 0_2|d^{\dagger}_{j
\uparrow}|0_1\rangle$.  An interesting question is the effect of cotunneling
through excited states of the double dot when the dot ring-coupling $|t|$
exceeds the many-body level spacing in the double dot $J$.  In order to 
address this question, we have studied the closed system of a double quantum
dot embedded in an AB ring numerically using the Lanczos technique\cite{dagotto}
(see Fig.\ \ref{ires.lanczos}).  The 
Hilbert space was truncated by discretizing the ring (8 sites were used).
In order to distinguish the contributions from tunneling through the  $S_z =0$
ground state and the $S_z=1$ excited state of the double dot, the total number 
of electrons in 
the system was chosen to be odd (in this case 5). Thus if the total
number of up-spin electrons $N_{\uparrow}$
is even, the total number of down-spin 
electrons $N_{\downarrow}$ must be odd, and vice versa.\cite{spin.z}  
In the weak-coupling limit  $|t| \ll J$,
where a single level of the double dot contributes to the resonant current,
the spin of the tunneling electron is well defined, and is
\begin{equation}
\hbar \sigma/2= \langle 0_N|S_z|0_N\rangle - \langle 0_{N-1} |S_z|0_{N-1}
\rangle.
\label{s.tunnel}
\end{equation}
Fig.\ \ref{ires.lanczos} shows the persistent current at $\phi=\pi/2$
as a function of the gate voltage $Q_g$ in the vicinity of the first Coulomb
blockade doublet centered near $Q_g=e/2$.  The doublet splitting is here
enhanced due to the finite level spacing in the ring.
For $\Delta>0$, the ground state of the coupled dot-ring
system generally has $N_{\downarrow}=N_{\uparrow}+1$ (in this case
$N_{\downarrow}=3$ and $N_{\uparrow}=2$). 
The first electron
to enter the double dot as $Q_g$ is increased from zero enters the lowest 
single-electron eigenstate of the double dot, and thus has $\sigma=-1$.
Since $N_{\downarrow}$ is odd, the resonant current is {\em
diamagnetic} due to the parity effect.\cite{mbcs}  The second electron
to enter the double dot goes into the state $|0_2\rangle$ and thus has 
$\sigma=+1$.  Since $N_{\uparrow}$ is even, the resonant current is thus
{\em paramagnetic}\cite{mbcs}
 (see solid curve in Fig.\ \ref{ires.lanczos}).  The height
of the second peak is reduced compared to that of the first, but by a smaller
factor than for the conductance [c.f., Eqs.\ (\ref{conductance}) and
(\ref{i.res})].  However,
there is also a contribution to the persistent current due to cotunneling 
through the first excited state of the double dot, which is higher in energy
by $J$ than $|0_2\rangle$.  This state has $\sigma=-1$, and thus leads
to a diamagnetic contribution to $I(\phi)$.  This state couples more
strongly to the leads, but is suppressed by a large energy denominator when
$|t| \ll J$.  As $|t|$ is increased (dotted and dashed curves in Fig.\
\ref{ires.lanczos}), the cotunneling contribution becomes increasingly 
important, and there is a crossover from a small paramagnetic peak to
a larger diamagnetic peak for $|t| > J$.  Fig.\ \ref{ires.lanczos}
clearly shows that the sign of the persistent current induced by tunneling
through a 1D structure may be used to characterize the spin quantum numbers
of the ground state and low-lying excited states of such a system.

\section{1D Array of Quantum Dots}

Let us next consider tunneling through larger arrays of quantum dots.  For
$N_d > 2$, the $N$-body ground states of Eq.\ (\ref{hubham}) were obtained
by the Lanczos technique,\cite{dagotto}
 and the conductance was calculated using Eq.\ (\ref{conductance}).
At $T=0$ and in the absence of inelastic scattering, the
conductance peaks all have height $e^2/h$ in the absence of disorder,
since in that case $\Gamma_N^R=\Gamma_N^L=\Gamma_N/2$.  Inelastic scattering
leads to additional broadening of the conductance peaks, and suppression of
the resonant conductance below $e^2/h$.  Disorder also leads to a suppression
of the $T=0$ resonant conductance below $e^2/h$ due to the breaking of 
left-right symmetry $\Gamma_N^R\neq\Gamma_N^L$.  In the following, we 
concentrate on the thermally broadened resonance regime, where the 
peak heights of the conductance resonances depend most strongly on the 
conductance matrix elements $\Gamma_N^L$ and $\Gamma_N^R$.  For $k_B T\gg
\Gamma_N$, Eq.\ (\ref{conductance}) simplifies to \cite{onedot,klimeck}
\begin{equation}
G=\frac{e^2}{h} \sum_N \frac{\Gamma_N^L\Gamma_N^R}{\Gamma_N^L+\Gamma_N^R}
[-f'(\mu-E_N^0+E_{N-1}^0)].
\label{g.therm}
\end{equation}

\begin{figure}
\vbox to 17cm {\vss\hbox to 17cm
 {\hss\
   {\includegraphics{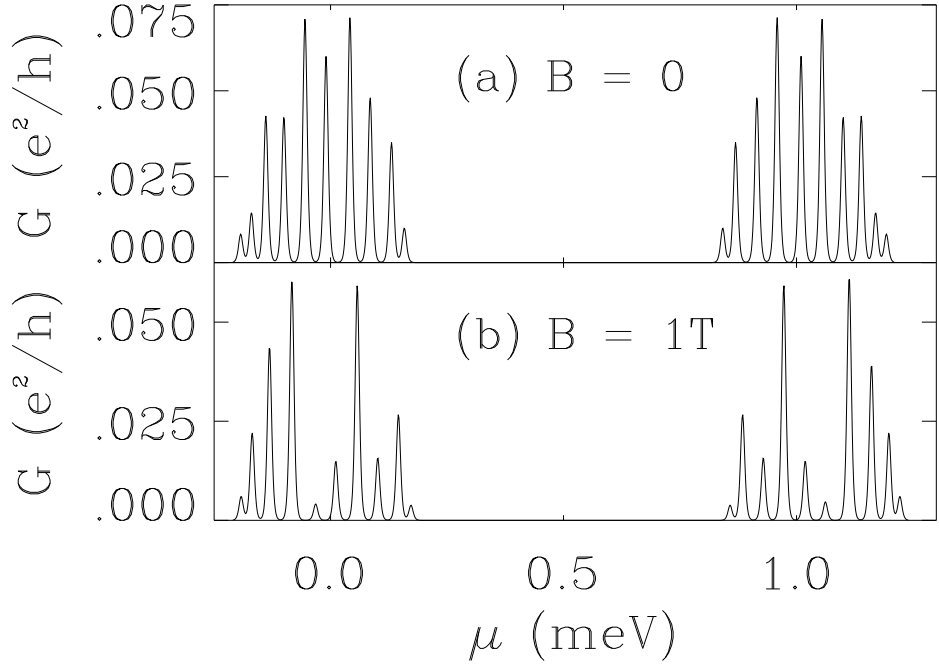} }
  \hss}
}
\caption{Conductance vs.\ chemical potential $\mu$ through a linear
array of 10 GaAs quantum dots with one spin-1/2 orbital per dot.
$e^2/C_g=1\mbox{meV}$, $C=0$,
$t=0.1\mbox{meV}$, and $T=35\mbox{mK}$.  Splitting of the two Coulomb
blockade peaks into minibands is driven by $t$.  The suppression of
the 5th peak in (b) is the result of a density-dependent
SPT.}
\label{fig1}
\end{figure}

Fig.\ \ref{fig1} shows the conductance through a linear array of 10 quantum
dots with $C =0$ 
as a function of the chemical potential $\mu$ in the leads, whose value
relative to the energy of the array is controlled by the gate voltages.
The two Coulomb blockade peaks in Fig.\ \ref{fig1}  are
split into multiplets of 10 by interdot tunneling, as discussed in
Refs.\ \onlinecite{us,klimeck}  We refer to these multiplets as {\em
Hubbard minibands}.  The energy gap between
multiplets is caused by collective Coulomb blockade,\cite{us}
and is analogous to the energy gap in a Mott
insulator.\cite{andyandi}  The heights of the resonant conductance peaks in
Fig.\ \ref{fig1}(a) can be understood as follows:
Since the barriers to the leads are assumed to be large, the single-particle
wavefunctions of the array are like those of a particle in a one-dimensional
box.  The lowest eigenstate has a maximum in the center 
of the array and a long wavelength, hence a small amplitude on the end dots,
leading to a suppression of the 1st conductance peak.  Higher
energy single-particle states have shorter wavelengths, and hence larger
amplitudes on the end dots, leading to conductance peaks of increasing
height.  The suppression of the conductance peaks at the top of the 1st
miniband can be understood by an analogous argument in terms of many-body
eigenstates; the 10th electron which enters the array can be thought of as
filling a single hole in a Mott insulator, etc.

In Fig.\ \ref{fig1}(b), the spin-degeneracy of the quantum dot orbitals is
lifted by the Zeeman splitting.  There is a critical field $B_c$ above which
the system is spin-polarized [c.f., Eq.\ (\ref{bcrit})]
Because $B_c$ is a function of $n$, one can pass through this spin-polarization
transition (SPT) by varying $n$ at fixed $B$.  In Fig.\ \ref{fig1}(b), this
transition occurs between the 4th and 5th electrons added
to the array, consistent with the prediction of Eq.\ (\ref{bcrit}).
The effect of this transition on the conductance spectrum is dramatic:  
The first 4 electrons which enter the array have spin aligned with $B$ (up),
but the 5th electron enters with the opposite spin, and goes predominantly 
into the lowest single-particle eigenstate for down-spin electrons, which
couples only weakly to the leads, leading to a suppression of the 5th
resonant conductance peak by over an order of magnitude.  It should be 
emphasized that the heights of the conductance peaks change {\em
discontinuously} as a function of $B$ each time there is a spin-flip.

Splitting of the Coulomb blockade peaks due to interdot coupling 
and suppression of the conductance peaks at the miniband edges 
have recently been observed experimentally
by Waugh {\it et al.} \cite{waugh}
However, it has been pointed out \cite{waugh} that both effects can also be
accounted for by a model \cite{classicaldots} 
of capacitively coupled dots with completely
{\em incoherent} interdot tunneling.  It is therefore of
interest to consider the effects of interdot capacitive coupling in the
regime of {\em coherent} interdot transport.  
A nonzero interdot capacitance $C$
introduces long-range electron-electron interactions in Eq.\ (\ref{hubham})
and decreases the intradot charging energy $U$.
Fig.\ \ref{chi_s} shows the spin susceptibility $\chi_s$
for $C/C_g=1/2$ in linear 
arrays with 8 electrons on 12 dots and 10 electrons
on 10 dots.  The $n$-dependence of $B_c$ in Fig.\ \ref{chi_s}
is qualitatively similar to that in a system with intradot interactions only,
but the values of $B_c$ are roughly twice
those of a system with $C=0$.
Note the rapid growth of $\chi_s$ as $B \rightarrow B_c$.
In an infinite array, $\chi_s$ is expected to diverge as $B \rightarrow B_c$
because the system undergoes a second order quantum phase transition.
\cite{crit}  The SPT predicted to occur in an array of coupled
quantum dots is in contrast to that observed in a single quantum dot,
\cite{klein} where the critical point occurs for minimum total spin.  

\begin{figure}
\vbox to 17cm {\vss\hbox to 17cm
 {\hss\
   {\includegraphics{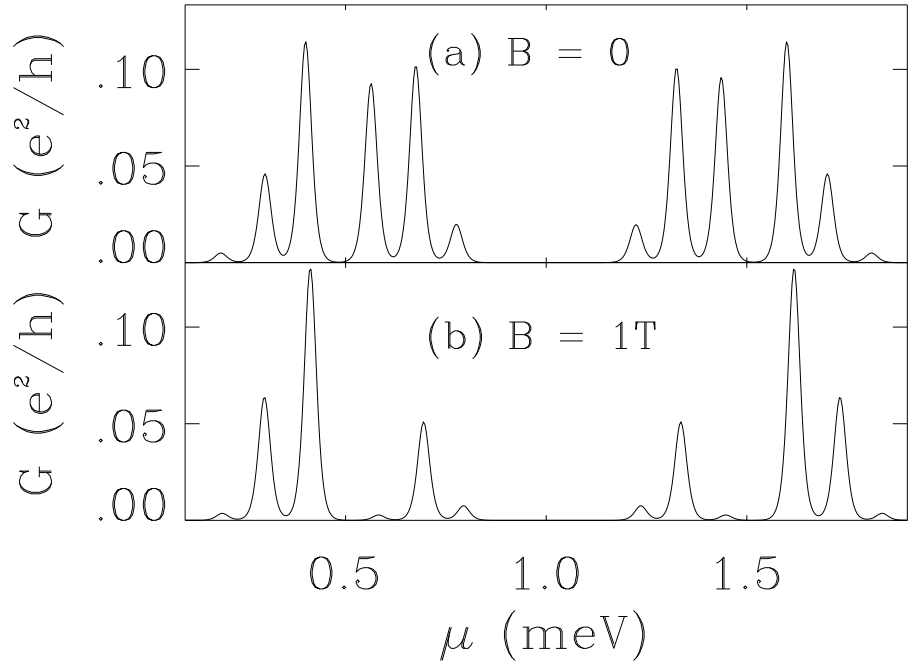} }
  \hss}
}
\caption{Conductance vs.\ chemical potential $\mu$ through a linear
array of 6 GaAs quantum dots with one spin-1/2 orbital per dot.
$e^2/C_g=1\mbox{meV}$, $C/C_g=0.5$,
$\bar{t}=.05\mbox{meV}$, $T=120\mbox{mK}$.  Disorder $\delta t/\bar{t}
\sim 1$ ($t_{i\uparrow}=t_{i\downarrow}$) is present in the hopping
matrix elements.  The splitting of the Coulomb blockade peaks into multiplets
is dominated by $C$; however, the effect of $B$ is similar to that in
Fig.\ 1.}
\label{fig2}
\end{figure}

Disorder introduces a length scale which cuts off the critical behavior
as $B \rightarrow B_c$.  However, as shown in Fig.\ \ref{fig2},
where disorder $\delta t \sim t$
has been included in the hopping matrix elements, the SPT
has a clear signature in the magnetotransport even in a
strongly disordered system.
In Fig.\ \ref{fig2}, the peak splitting due to
capacitive coupling is roughly ten times that due to interdot tunneling, so
that the peak positions are within $\sim 10\%$ of those predicted 
by a classical charging model.\cite{classicaldots}  However, the dramatic
dependence of peak heights on magnetic field---the 4th conductance peak in
Fig.\ \ref{fig2}(b) is suppressed by a factor of 32 compared to its $B=0$
value due to the 
density-dependent SPT described above---can not be accounted 
for in a model which neglects coherent interdot
tunneling.  This effect should be observable provided $g\mu_{B} B_c > 
\mbox{max}(k_B T,\hbar/\tau_i)$, where $\tau_i$ is the inelastic scattering
time.  We believe that this striking magnetotransport effect is the clearest
possible signature of a coherent molecular wavefunction in an array of
quantum dots.

\begin{figure}
\vbox to 17cm {\vss\hbox to 17cm
 {\hss\
   {\includegraphics{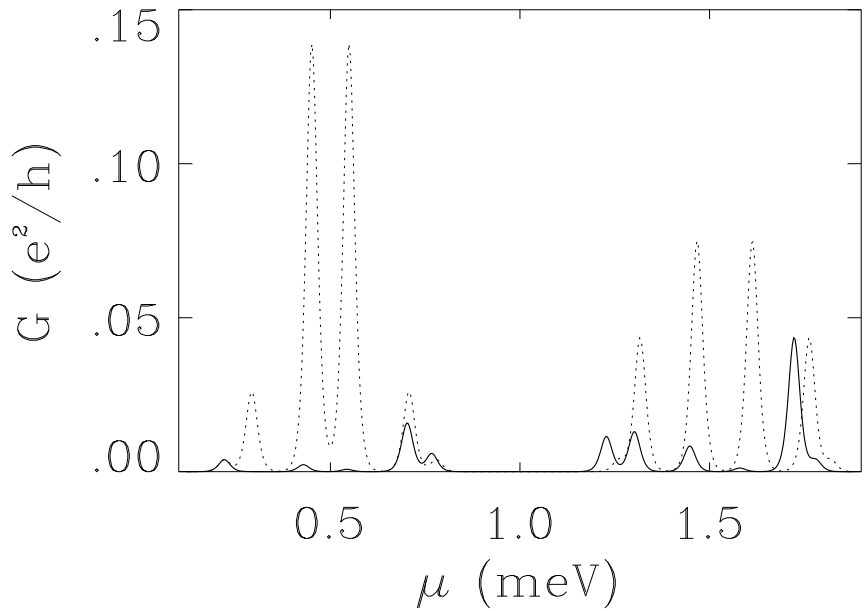} }
  \hss}
}
\caption{Conductance vs.\ chemical potential $\mu$ through a linear
array of 6 GaAs quantum dots with one spin-1/2 orbital per dot.
$e^2/C_g=1\mbox{meV}$, $C/C_g=0.5$,
$\bar{t}=.05\mbox{meV}$, $T=120\mbox{mK}$.  Spin-dependent
disorder $\delta t/\bar{t}
\sim 1$ ($t_{i\uparrow}\ne t_{i\downarrow}$) is 
included in the hopping matrix elements.  Solid curve: $B=0$; dotted curve:
$B=1.3 \mbox{T}$.  At 1.3T, the 2nd conductance peak is enhanced by a factor
of 1600.}
\label{fig3}
\end{figure}

Fig.\ \ref{fig3} shows the conductance spectrum for an array of 6 quantum
dots with the same parameters as in Fig.\ \ref{fig2}, but with
spin-dependent disorder in the hopping matrix elements, as could be
introduced by magnetic impurities.  Several conductance peaks at $B=0$
(solid curve) are strongly suppressed due to a many-body enhancement of
localization.  This effect arises because repulsive on-site interactions
enhance spin-density wave correlations, which are pinned by the
spin-dependent disorder.\cite{tands}  At $B=1.3 \mbox{T}$ (dotted curve)
the system is above $B_c$ and is spin-polarized, circumventing this effect.
The second conductance peak is enhanced by a factor of 1600 at 1.3T compared to
its size at $B=0$ (not visible on this scale).  This {\em giant
magnetoconductance} effect is a many-body effect
intrinsic to the regime of coherent interdot transport.

\begin{figure}
\vbox to 17cm {\vss\hbox to 17cm
 {\hss\
   {\includegraphics{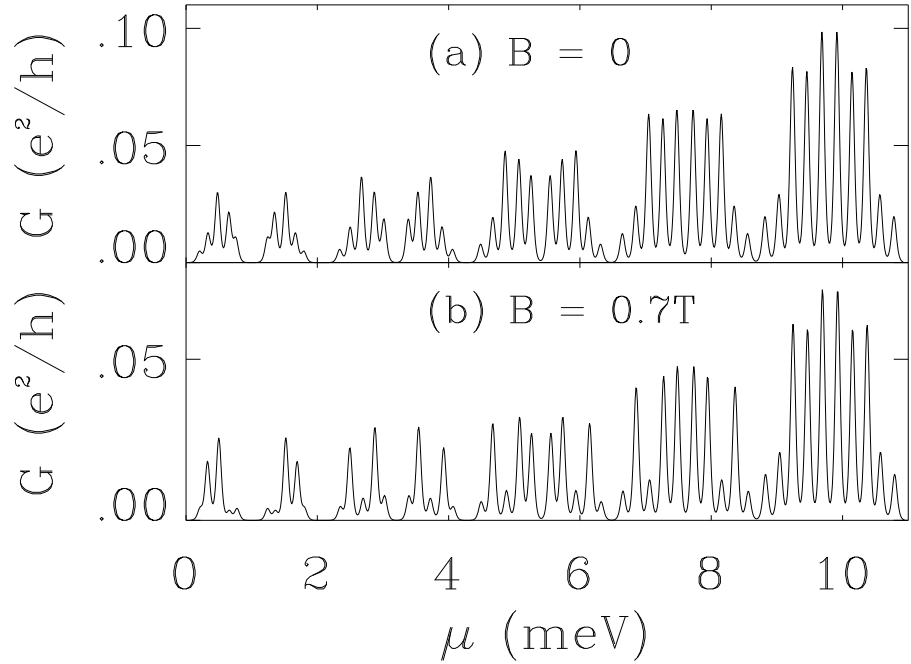} }
  \hss}
}
\caption{Conductance vs.\ chemical potential
through a linear array of 5 GaAs quantum
dots with 5 spin-1/2 orbitals per dot.  
$e^2/C_g=1\,\mbox{meV}$, $\Delta = 0.2\,\mbox{meV}$, $T=.29\mbox{K}$,
$C^{(n)}/C_g=2^{n-1}$, and $t_{n} = 0.05\,\mbox{meV}
(1.05)^{n}$ ($n=0,\ldots,4$).
The energy gap between Hubbard minibands is not
resolved for $\mu > 9\,\mbox{meV}$ (breakdown of CCB). 
Note the quenching of magnetoconductance effects in the ballistic regime.}
\label{fig.mit.longrange}
\end{figure}

Another interesting phenomenon stemming from the competition between coherent
interdot tunneling and charging effects is the Mott-Hubbard metal-insulator
transition (MH-MIT), which occurs when collective Coulomb blockade (CCB) 
\cite{us} is destroyed
due to strong interdot coupling.  For GaAs quantum dots larger than 
about 100nm in diameter, we find that this transition is caused by the 
divergence of the effective interdot capacitance, similar to
the breakdown of Coulomb blockade in a single quantum dot.\cite{cap2}
Within the framework of the scaling theory of the MH-MIT,
\cite{andyandi} one
expects a crossover from CCB to ballistic transport in a finite array of
quantum dots when the correlation
length $\xi$ in the CCB phase significantly exceeds the linear dimension $L$
of the array.  Fig.\ \ref{fig.mit.longrange} 
shows the conductance spectrum for 5 quantum dots
with 5 spin-1/2 orbitals per dot. 
The divergence of the effective interdot capacitance as the interdot
barriers become transparent is simulated by setting
$C^{(n)}/C_g = 2^{n-1}$, $n=0,\ldots,4$.  In Fig.\ \ref{fig.mit.longrange},
minibands arising from each orbital are split symmetrically into multiplets
of 5 peaks by CCB, with the center to center spacing between
multiplets equal to $e^2/C_g$, while the energy gap
between minibands corresponds to the band gap $\sim \, \Delta$
enhanced by charging effects.  
The CCB energy gap is evident in the first 3 minibands, 
but is not resolvable for the
higher orbitals ($C/C_g \geq 4$),
although there is still a slight suppression of the 
conductance peaks near the center of the 4th miniband.
Comparison of the compressibility of the system 
to a universal scaling function for the MH-MIT calculated by the method of
Ref.\ \onlinecite{andyandi} indicates $\xi/L \sim 10^3$ for $C/C_g = 8$,
so that the transport in the 5th miniband is effectively ballistic.
The peak spacing within a miniband
saturates at $e^2/LC_g$ (plus quantum corrections $\sim t/L$)
in the ballistic phase 
because the array behaves like one large capacitor, as observed experimentally
in Ref.\ \onlinecite{waugh}. 
Fig.\ \ref{fig.mit.longrange}(b) shows the effects of a magnetic field on the
conductance spectrum:  a sequence of SPTs is
evident in the different minibands, with $B_c$ an
increasing function of $C/C_g$, leading to quenching of magnetoconductance
effects in the ballistic regime.

A finite-size scaling analysis of the compressibility indicates that the
MH-MIT probably occurs at $C/Cg =\infty$ in an infinite array of quantum
dots, when the interdot barriers become transparent to
one transmission mode.\cite{smalldots}  

\section{2D array of quantum dots}

\begin{figure}
 \vbox to 8.5cm {\vss\hbox to 6cm
 {\hss\
{\includegraphics{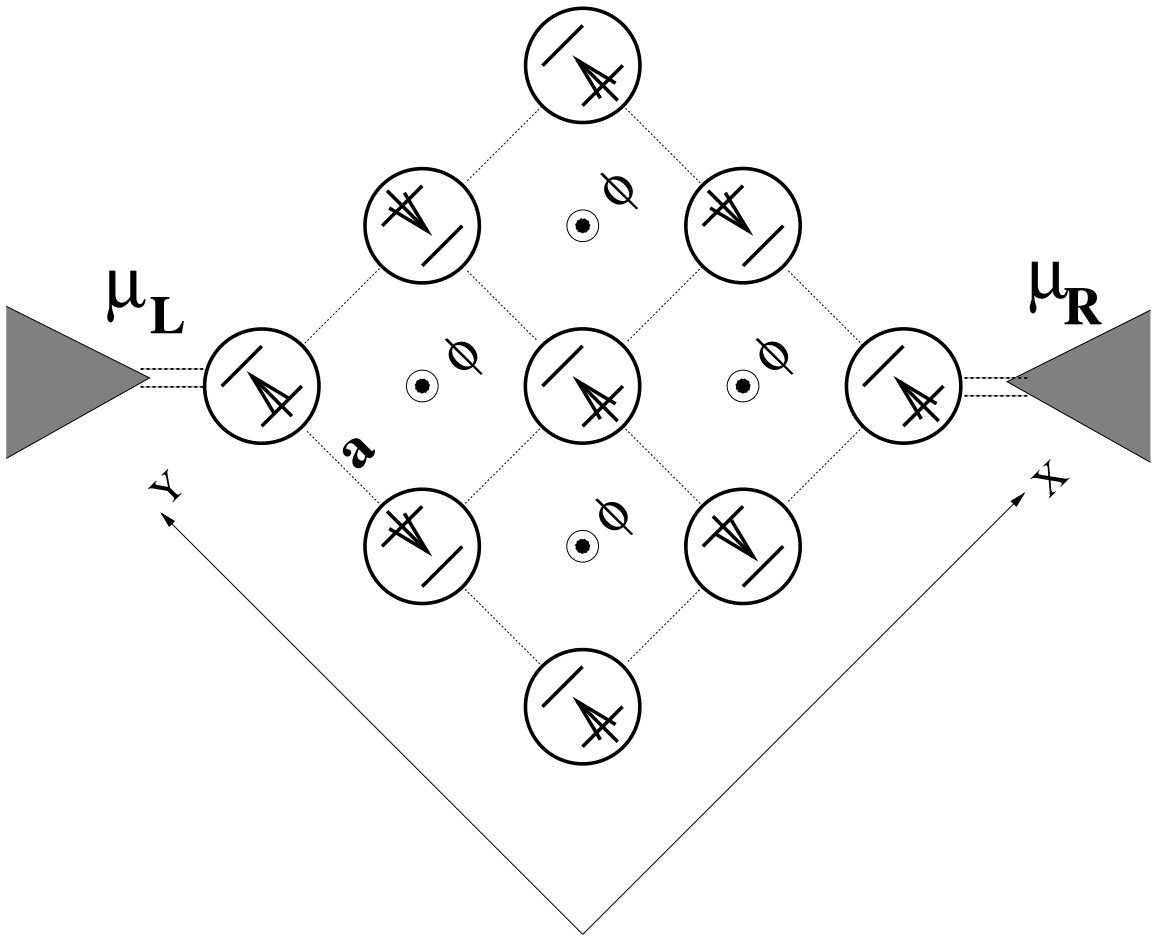}
   }
  \hss}
 }
\vspace{6cm}
\caption{Schematic diagram of
a $3 \times 3$ array of quantum dots.}
\label{10.fig}
\end{figure}

Finally, we briefly consider coherent tunneling through a 
2D quantum dot array (Fig.\ \ref{10.fig}) to investigate whether 
the many-body {\em giant magnetoconductance} effect discussed 
in Sec.\ V arises in two dimensions as well.
We use Eq. (\ref{g.therm}) to calculate the linear tunneling conductance
 through a 2D 
$3 \times 3$ quantum dot 
array consisting of $N_{d}=9$ quantum dots. The corner dots in the 
array are  weakly coupled to electron reservoirs as shown in Fig. \ref{10.fig}.
The Hamiltonian of the array is the same as that 
given by Eq. (\ref{hubham}), where the 
second term 
is now modified to incorporate the nearest neighbor tunneling in the 2D array: 
the tunneling amplitudes connecting two nearest neighbor dots $i$ 
and $j$ in the array 
 are multiplied by the Peierls phase 
factors $e^{i \frac{e}{\hbar } \int_{ij} \vec{A} \cdot \vec{dl_{ij}}}$ with
$\vec{A}$ as  the magnetic   vector  potential.\cite{kotlyarsdsPCUR} 
We consider  a uniform flux $\phi=B a^{2}$
piercing each unit cell of
the array in Fig. \ref{10.fig}.
The magnetic
field $B$ enters through the tunneling amplitudes $t_{ij}$  
 and through the intradot single-particle field
dependence entering $\varepsilon_{i\sigma}$ in Eq. (\ref{site.energy}). 
The flux sensitive phase factors in Eq. (\ref{hubham}) lead to a
flux periodic modulation of the linear conductance with periodicity 
given by one fundamental flux unit $hc/e$. 
This flux dependence of the linear 
conductance and the associated ground state persistent current oscillations
have been discussed elsewhere.\cite{kotlyarsdsPCUR}
Here we follow our discussion in the previous section, concentrating 
on a fixed applied field, and 
focus on the magnetic field induced spin effect (i.e. the
 single-dot Zeeman  physics) on the linear conductance 
peak heights in the $3 \times 3$ array.

\begin{figure}
 \vbox to 8.5cm {\vss\hbox to 6cm
 {\hss\
{\includegraphics{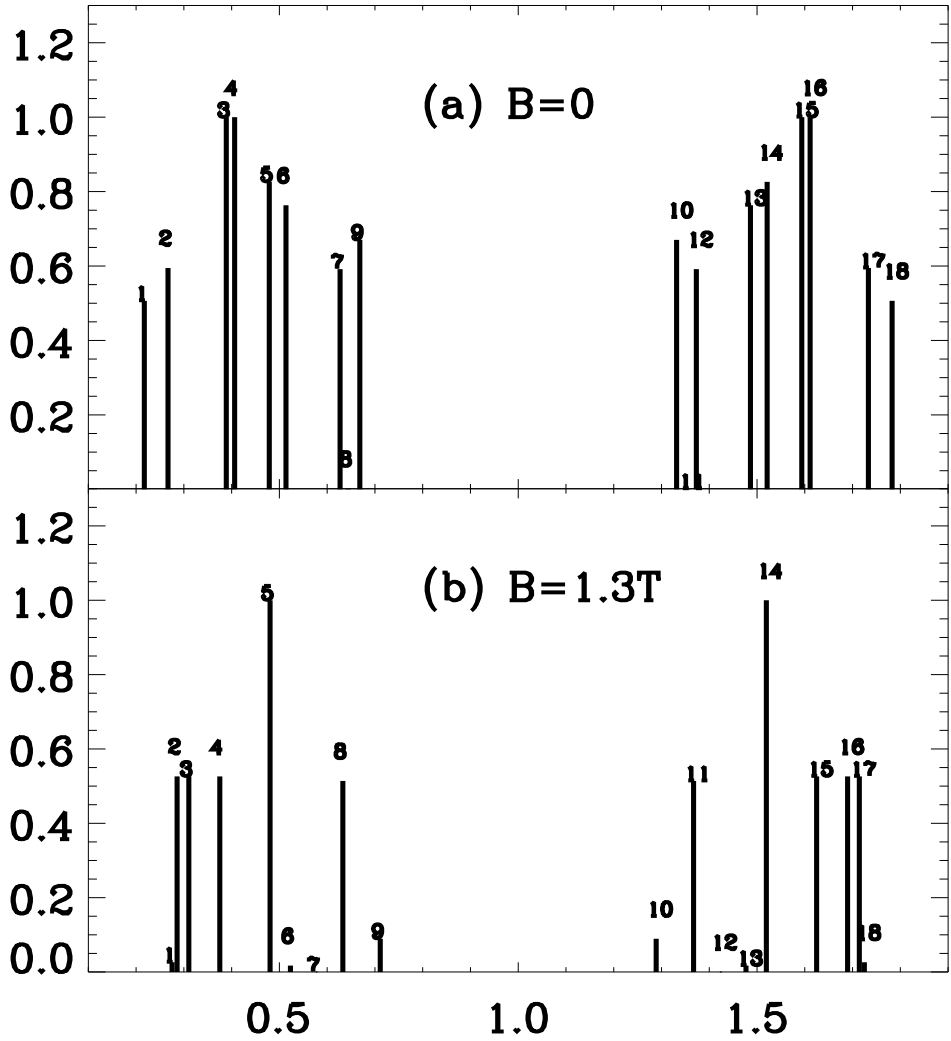}
   }
  \hss}
 }
\vspace{6cm}
\caption{The partial width
$\Gamma_{N}^{L}=\Gamma \sum_{\sigma}
| \langle 0_{N}|d_{1\sigma}^{\dagger}| 0_{N-1} \rangle |^{2}$
as a function of $N$ for the $3 \times 3$ array of quantum dots
with pure Hubbard interactions $U=10t$.
The widths are plotted normalized by the partial width $\Gamma_{4}^{L}$ in
(a) and  $\Gamma_{5}^{L}$ in (b). The $x$ coordinate of the $N$th peak is
given by $E_{N}^{0}-E_{N-1}^{0}$. Each peak is labeled by $N$. The peak
structure is discussed in the text.}    
\label{11.fig}
\end{figure}

\begin{figure}
 \vbox to 8.5cm {\vss\hbox to 6cm
 {\hss\
{\includegraphics{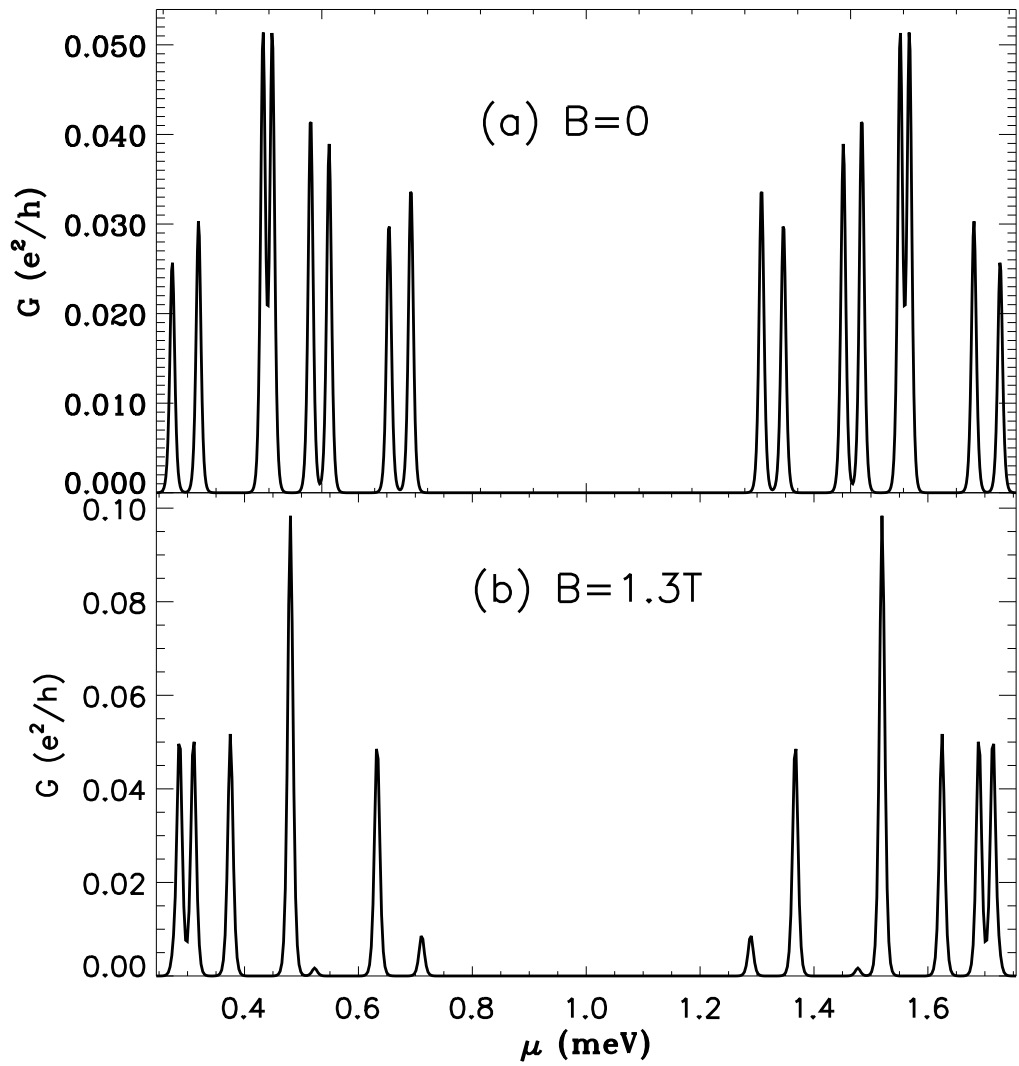}
   }
  \hss}
 }
\vspace{6cm}
\caption{Conductance vs.  chemical potential $\mu$
through the $3 \times 3$ array of 
quantum dots. The same parameters used as in Fig.\ 6.
 The peak structure is discussed in the text.}
\label{12.fig}
\end{figure}

The partial width $\Gamma_{N}^{L}$ of the $N$th resonance is plotted as 
a function of $N$ in Fig.\ \ref{11.fig} (a) and (b) for 
the applied magnetic field $B=0$ and $B=1.3$T 
in the array, respectively. The corresponding linear conductance 
in the $3 \times 3$ array is shown in Fig.\ \ref{12.fig}. The peak splittings 
in the linear conductance in the $3 \times 3$ array are not 
distributed uniformly, but the shape of the envelope function for 
conductance peak heights is similar to that in 1D chains (e.g. compare 
Figs.\ \ref{fig1} and 
\ref{12.fig}). This envelope function is peaked at quarter filling 
in the Hubbard model we use here. In the strongly correlated Hubbard array, 
the overlap matrix element is approximately given by 
$\Gamma_{N}^{L} \sim [1-P_{N-1}(1)] P_{N}(1)$ with 
$ P_{N}(1) = \sum_{\sigma} 
\langle 0_{N}|d_{1\sigma}^{\dagger}d_{1\sigma}| 0_{N} \rangle$ being the 
probability to find the corner dot occupied. For $U/t=10$, the maximum 
and minimum $P_{N}(1)$ are approximately 1 and 0, leading to a peak
in the linear conductance at quarter filling (and also at three quarter 
filling due to particle-hole symmetry) in Hubbard arrays. 
The  addition spectra shown  in the conductance vs.\ chemical potential 
plots of Figs.\ \ref{11.fig} and 
\ref{12.fig} correspond to a sequence of ground 
states that are characterized by the number of electrons $N$, the 
total spin $S$, and the component $S_{z}$ of the total spin along 
the quantization axis. The ground state $(N,S,S_{z})$ sequence for 
$B=0$ in the array is 
$(1,1/2,1/2)\rightarrow (2,0,0)\rightarrow (3,3/2,1/2) 
\rightarrow (4,1,1)\rightarrow (5,1/2,1/2)\rightarrow (6,0,0)
\rightarrow (7,1/2,1/2)\rightarrow (8,0,0)\rightarrow (9,1/2,1/2)$. 
In the $B=1.3$T array the similar sequence of the ground states is given by
$(1,1/2,1/2)\rightarrow (2,1,1)\rightarrow (3,3/2,3/2)
\rightarrow (4,2,2)\rightarrow (5,5/2,5/2)\rightarrow (6,2,2)
\rightarrow (7,1/2,1/2)\rightarrow (8,1,1)\rightarrow (9,1/2,1/2)$. 
It can be seen from the latter sequence that the 
$6$th linear conductance peak at $B=1.3$ is suppressed 
(by a factor of 23)  due 
to the spin polarization transition discussed in the previous 
section for 1D arrays. It can also be seen that the transition 
from the $6$-electron to $7$-electron ground state is forbidden at 
$B=1.3T$ and therefore the $7$th peak is absent in Figs.\ \ref{11.fig}(b) and 
\ref{12.fig}(b). This is an example of the so-called ``spin-blockade'' 
phenomenon.\cite{sblockade} The $8$th peak at $B=0$ in the array is 
present in the conductance 
and partial width traces in  Figs.\ \ref{11.fig}(a) and 
\ref{12.fig}(a), although it is suppressed by a factor of approximately 100. 
Finally, for $N=1$ 
the charge density on the corner dots is 10 times smaller at $B=1.3$T, leading 
to the suppression of the first peak at $B=1.3$T  from its $B=0$ value in 
Fig.\ \ref{11.fig}.  

We conclude that the spin polarization transition discussed in Sec.\ V 
for the case of the 1D array
leads to a similar suppression 
of the linear conductance peak heights in 2D arrays of 
quantum dots. 

\section{Conclusions}

We have shown that the formation of artificial molecular
bonds due to interdot superexchange 
can drastically modify the low-temperature transport through
coupled quantum dots.  The resulting interdot spin-density-wave (SDW)
 correlations
are strongly pinned by magnetic disorder, leading to a suppression of transport.
These SDW correlations are destroyed in an applied magnetic field large enough
to polarize all the electron spins, leading to a marked 
increase of the conductance
at the spin-polarization transition (SPT).  For a double quantum dot, this
leads to a magnetoresistance proportional to $(g\mu_B h/e^2)U^3/t^4$, 
where $U$ is the 
charging energy of a quantum dot and $t$ is the interdot hopping matrix 
element.  Since $U$ is typically at least an order of magnitude greater
than $t$, we have termed this effect {\em giant magnetoresistance}.
For larger 1D arrays of quantum dots, the magnetoresistance was found to be
proportional to $(g\mu_B h/e^2) U^{2N-1}/t^{2N}$ at the $N$-electron resonance
when $N$ is even, while saturating to a smaller, $N$-independent value for
$N$ odd.  This $U$-dependence reflects the probability of an electron to
tunnel all the way through the system while leaving the pinned, N\'eel ordered 
spin configuration of the ground state undisturbed.
The giant magnetoresistance effect proposed here for coupled quantum dots
is expected to be quite generic in quasi-one-dimensional
systems with magnetic disorder.

In addition to the giant magnetoresistance effect predicted for 1D arrays
of quantum dots with
magnetic disorder, the SPT was shown to lead to large magnetoresistance 
effects in non-disordered 1D and 2D systems and in systems with charge disorder.
These many-body effects are likely to provide a fruitful area of research in 
the next generation of coupled semiconductor quantum dot systems.

\begin{center}
{\bf ACKNOWLEDGMENTS}
\end{center}

This work was supported by the U.S.\ Office of Naval Research.

\pagebreak

\pagebreak

\pagebreak

\pagebreak

\pagebreak

\pagebreak

\pagebreak

\pagebreak

\pagebreak

\pagebreak

\pagebreak

\pagebreak

\begin{references}
\bibitem{noandreev} Eq.\ (\ref{multi}) is applicable to tunneling through
superconducting systems in which the charging energy exceeds the 
superconducting gap, so that Andreev reflection is energetically prohibited.
\bibitem{spin.z} Note that 
$N_{\uparrow}$ and $N_{\downarrow}$ commute with the Hamiltonian, Eqs.\
(\ref{hubham}) and (\ref{ham}), although the total spin $S^2$ does not, due
to the spin-dependent disorder, Eq.\ (\ref{site.energy}).
\bibitem{smalldots} In contrast, for coupled GaAs quantum dots smaller than
about 100nm in diameter, the breakdown of CCB is expected to occur \cite{us} 
for $t \sim U$, when the transmission through the interdot barriers is less
than unity.
\end{references}
\end{document}